\documentclass[9pt,twocolumn,twoside]{opticajnl}
\journal{opticajournal} % use for journal or Optica Open submissions
\setboolean{shortarticle}{true}
\usepackage{lineno}
\usepackage{graphicx}
\usepackage{comment}
\usepackage{subcaption}
\title{High fidelity preservation of photonic hyperentanglement in a free-space optical delay line}

\author[1,*]{Yu Guo}
\author[1,2,3]{Arya Chowdhury}
\author[2] {Pranay Tiwari}
\author[1] {Jia Boon Chin}
\author[1] {Anindya Banerji}
\author[1,2]{Alexander Ling}

\affil[1]{Centre for Quantum Technologies, National University of Singapore, Block S15, Science Drive 2, Singapore 117543}
\affil[2]{Department of Physics, National University of Singapore, 2 Science Drive 3, Singapore 117551}
\affil[3]{Thales Research and Technology, 12 Ayer Rajah Crescent, Singapore 139941}

\affil[*]{guoyu04@u.nus.edu}

\begin{abstract}
Photonic hyperentanglement enables increased information capacity and enhanced functionality for quantum communication and networking. However, synchronization of hyperentangled photon pairs requires maintaining correlations simultaneously across multiple degrees of freedom (DOFs). The preservation of polarization and energy-time entanglement in hyperentangled photon pairs is demonstrated using a free-space optical delay line based on nested Herriott cells. After a delay of 647.0(6) ns, a two-photon interference visibility of 93.9(3)\% is observed in the energy-time DOF, while a CHSH parameter of 2.758(5) is obtained in the polarization DOF. These results confirm that entanglement correlations in both DOFs are preserved after propagation through the delay line. They demonstrate that free-space optical delay lines are compatible with complex photonic quantum states and provide a promising route toward delay-based memories for synchronization and multiplexing in quantum networks.
\end{abstract}

\setboolean{displaycopyright}{false} % Do not include copyright or licensing information in submission.

\begin{document}

\maketitle

%\section{Stucture}

Hyperentanglement refers to simultaneous entanglement across multiple degrees of freedom (DOFs) of a quantum system \cite{Deng2017SciBull,Barreiro2005PRL}. Typical implementations for photon pairs combine polarization with energy-time or orbital angular momentum, enabling access to higher dimensional correlations beyond single-DOF states \cite{Suo2015OE,Liu2014PRL,prilmuller2018hyperentanglement,li2015hyperentanglement}. Such states enhance the information capacity per photon pair and expand the accessible state space for photonic quantum protocols \cite{Deng2017SciBull}.

%These advantages make hyperentanglement a powerful resource for quantum networks, enabling multiplexing, high-dimensional encoding, and more efficient entanglement distribution. Such capabilities support scalable networking tasks and improved resilience to channel noise \cite{Deng2017SciBull}.
The enlarged state space makes hyperentanglement a useful resource for quantum networks, where multiple DOFs can support multiplexing, high dimensional encoding \cite{cozzolino2019high}, and more efficient entanglement distribution \cite{steinlechner2017distribution}. Such capabilities are relevant for scalable networking tasks and improved resilience to channel noise \cite{Deng2017SciBull}.

Temporal synchronization is a general requirement for photonic quantum networks, where probabilistic entanglement generation and routing impose stringent timing constraints \cite{Briegel1998PRL,Sangouard2011RMP}. For hyperentangled states, a delay or storage device must preserve the state quality in each individual DOF without introducing coupling between them. Channel imperfections such as dispersion and polarization-dependent effects may degrade the correlations in individual DOFs or introduce cross-correlations \cite{Simon2010EJP}. Demonstrating preservation of both polarization and energy-time correlations after a well-defined delay is therefore essential for establishing hyperentanglement as a usable network resource.

The storage of photonic hyperentanglement is an active field of research. In particular, solid-state quantum memories have demonstrated storage of polarization and energy-time hyperentanglement, as well as multi-DOF storage involving polarization and orbital angular momentum \cite{Tiranov2015Optica,Nicolas2014NatPhoton}. These solid-state memories, however, require cryogenic cooling and can have limited retrieval efficiencies and acceptance bandwidths, which constrain their use with broadband multi-DOF states~\cite{Zhou2015PRL}.

A free-space optical delay line (FSODL) provides an alternative route for photonic storage by introducing a well-defined propagation delay that can be reused in a delay-based memory \cite{arnold2024all,wang2024low}. A recent demonstration of an FSODL based on nested Herriott cells showed high transmission efficiency, ultra broad bandwidth, and a large time-bandwidth product at room temperature \cite{guo2026highly}. Its all-reflective geometry minimizes chromatic dispersion and supports stable polarization propagation, making this platform attractive for broadband photonic quantum states. However, each mirror reflection may introduce small polarization-dependent phase shifts and losses. The cumulative effects over many reflections could degrade the entanglement in each DOF or introduce coupling between the DOFs. Verifying the preservation of hyperentanglement after many reflections is therefore essential for establishing the FSODL as a viable platform for synchronizing multi-DOF photonic states.

%This work reports the preservation of both polarization and energy-time entanglement in hyperentangled photon pairs after propagation through a nested Herriott cell based FSODL. High visibility energy-time interference and strong polarization correlations are observed after the delay, confirming that correlations which define entanglement in each DOF are preserved. These results suggest that FSODLs are compatible with complex photonic quantum states relevant to synchronization and multiplexing in quantum networks.
\begin{figure*}[ht]
\centering
\includegraphics[width=0.95\textwidth,height=7cm]{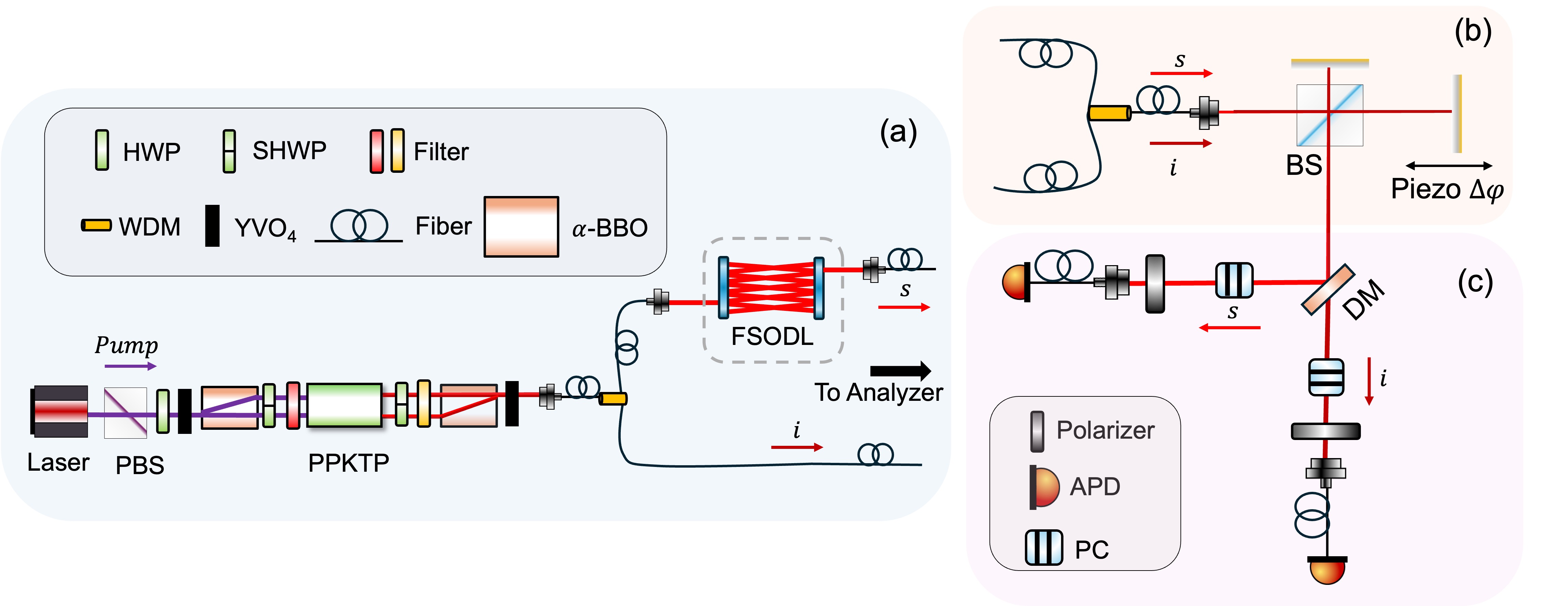}
\caption{Experimental setup for generation, propagation, and characterization of hyperentangled photon pairs in a free-space optical delay line (FSODL). 
(a) Hyperentangled photon pairs are generated via spontaneous parametric down-conversion (SPDC) in a PPKTP crystal. The signal photon propagates through the FSODL, while the idler photon serves as a reference. 
(b) Energy-time entanglement is characterized using an unbalanced interferometer with a piezo-controlled phase $\Delta \varphi$. 
(c) Polarization correlations are measured using polarization analyzers for the signal and idler photons.
PBS: polarization beam splitter; BS: beam splitter; PPKTP: periodically poled potassium titanyl phosphate crystal; YVO$_{4}$: yttrium orthovanadate crystal; $\alpha$-BBO: alpha-barium borate crystal; DM: dichroic mirror; HWP: half-wave plate; SHWP: segmented half-wave plate; WDM: wavelength division multiplexer; APD: silicon avalanche photodiode; PC: polarization compensation module implemented with a quarter-half-quarter wave plate set to compensate fiber-induced polarization rotation.}
\label{fig:Setup}
\end{figure*}

This work reports the preservation of both polarization and energy-time entanglement in hyperentangled photon pairs after propagation through a nested Herriott cell based FSODL. After the FSODL, the measured energy-time interference visibility is 93.9(3)\%. The polarization-state fidelity relative to the undelayed state is 99.5(4)\%. The measured CHSH parameter is 2.758(5). These results establish that FSODLs are compatible with complex photonic quantum states relevant to synchronization and multiplexing in quantum networks.

The hyperentangled photon pairs used in this work are generated through spontaneous parametric down-conversion (SPDC) in the experimental setup shown in Fig.~\ref{fig:Setup}(a). 
A continuous-wave pump laser at 405.2~nm is directed into a type-0 periodically poled KTP (PPKTP) crystal, generating signal and idler photons through SPDC at 780.0~nm and 842.6~nm, with full width at half maximum (FWHM) spectral bandwidths of 2.6~nm and 2.8~nm, respectively.
Under continuous-wave pumping, the long coherence time of the pump renders the photon-pair emission time uncertain, while energy conservation imposes strong frequency anti-correlations between the signal and idler photons. When measured using an unbalanced interferometer \cite{franson1989bell}, two-photon interference is observed between the indistinguishable short-short and long-long alternatives contributing to the central coincidence peak. The corresponding effective time-bin state can be written as
\begin{equation}
|\Phi\rangle_{E}=\frac{1}{\sqrt{2}}\left(|S_{s}S_{i}\rangle+e^{i\varphi}|L_{s}L_{i}\rangle\right),
\end{equation}
where the subscripts $s$ and $i$ denote the signal and idler photons; $S$ and $L$ denote the short and long interferometer paths; and $\varphi$ is the relative phase between the two-photon amplitudes.

Polarization entanglement is generated using a linear displacement interferometer design  \cite{lohrmann2020broadband}, resulting in the polarization entangled state
\begin{equation}
|\Phi\rangle_p=\frac{1}{\sqrt{2}}\left(|H_sH_i\rangle+e^{i\theta}|V_sV_i\rangle\right),
\end{equation}
where $\theta$ is the relative phase between the $|H_sH_i\rangle$ and $|V_sV_i\rangle$ components.

The resulting two-photon state is hyperentangled in the polarization and energy-time DOFs, with the joint state given by
\begin{equation}
|\Phi\rangle = |\Phi\rangle_p \otimes |\Phi\rangle_E.
\end{equation}

The FSODL used in this work is illustrated in Fig.~\ref{fig:Setup}(a). It consists of two nested concave mirrors that guide photons through multiple reflections and introduce a well-defined propagation delay. The combination of free-space propagation and mirror reflections introduces negligible chromatic dispersion and polarization-dependent distortions \cite{guo2026highly,ou2023generalized}. These features are well-suited for preserving energy-time and polarization correlations in hyperentangled states. In the present configuration, the mirror coating is specified to support operation across a wavelength range from 750 nm to 810 nm, corresponding to a bandwidth of approximately 30 THz.

%The signal photon from the SPDC source is directed into the optical delay line, while the idler photon propagates through a reference path for coincidence measurements. In the present configuration, the delay line introduces a temporal delay of approximately $647~\mathrm{ns}$ for the signal photon. After propagation, the signal and idler photons are directed to measurement analyzers for characterization of energy-time correlations (Fig.~\ref{fig:Setup}(b)) and polarization (Fig.~\ref{fig:Setup}(c)) \cite{Tiranov2015Optica,prilmuller2018hyperentanglement,lu2023generation}.

The signal photon from the SPDC source is directed into the FSODL, while the idler photon propagates through a reference path for coincidence measurements. The signal and idler photons traverse equal 1 $m$ fiber sections, and the resulting dispersion-induced temporal broadening is negligible compared with the detector timing jitter. In the current setup, the delay line introduces a temporal delay of 647.0(6)~ns for the signal photon after 160 reflections, with the uncertainty limited by the coincidence timing resolution. The delayed photon is recoupled into a single mode fiber with an efficiency of 73.9(3)\% and is subsequently directed to the energy-time and polarization analysis setups shown in Figs.~\ref{fig:Setup}(b) and (c) \cite{Tiranov2015Optica,prilmuller2018hyperentanglement,lu2023generation}, respectively.

\begin{figure*}[ht]
\centering
\includegraphics[width=\textwidth,height = 4.5cm]{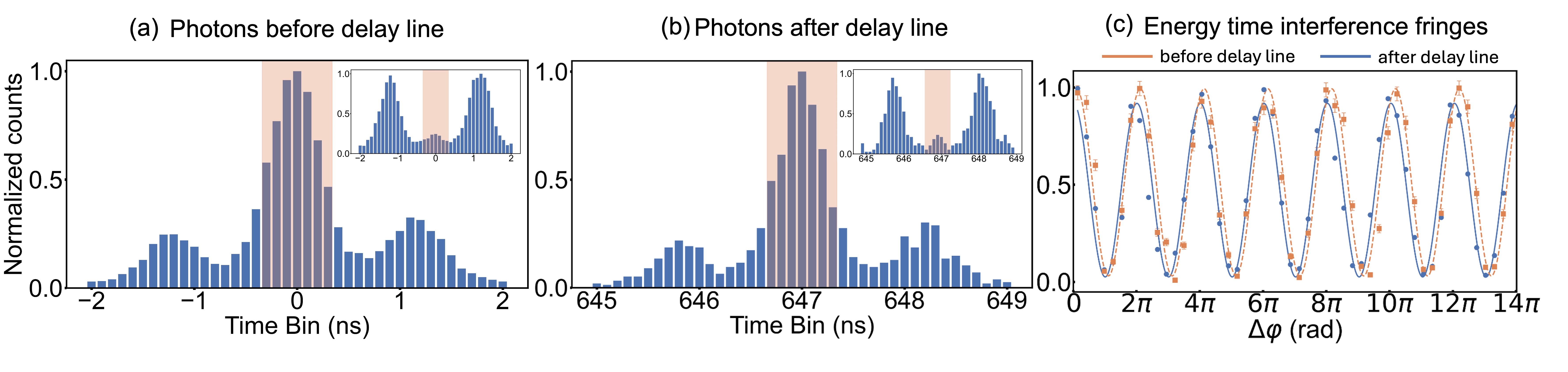}
\caption{Time-bin-resolved coincidence measurements demonstrating the preservation of energy-time entanglement through the FSODL.
(a) Coincidence histogram of photons before the delay line under constructive interference, centered at zero delay. The inset shows the corresponding histogram under destructive interference.
(b) Coincidence histogram of delayed photons after propagation through the delay line under constructive interference, centered at $647.0(6)~\mathrm{ns}$. The shaded regions indicate the 700~ps post-selection window used to extract coincidence counts for interference analysis. For each panel, the main histogram and the inset are normalized to the maximum central-peak count of the respective constructive-interference measurement.
(c) Normalized coincidence counts as a function of the two-photon phase shift $\Delta \varphi$ for photons measured before and after the delay line.}
\label{fig:EnergyHistogram}
\end{figure*}

For energy-time measurements, the signal and idler photons are sent into an unbalanced Michelson interferometer \cite{kwiat1990correlated}, as illustrated in Fig.~\ref{fig:Setup}(b). The path imbalance is chosen to exceed the single-photon coherence length while remaining within the two-photon coherence length, enabling interference between the indistinguishable short-short and long-long two-photon amplitudes. The relative phase is scanned by a piezo actuator attached to one of the interferometer mirrors. A mirror displacement $x$ introduces a round trip optical path change of $2x$, resulting in a two-photon phase shift
\begin{equation}
\Delta\varphi = 4\pi x\left(\frac{1}{\lambda_s}+\frac{1}{\lambda_i}\right),
\end{equation}
where $\lambda_s$ and $\lambda_i$ are the wavelengths of the signal and idler photons. Coincidence events are post-selected within the central time-bin to isolate the short-short and long-long contributions for the energy-time interference measurement.

\begin{figure}[t]
\centering
\includegraphics[width = 0.98\linewidth,height=7.8cm]{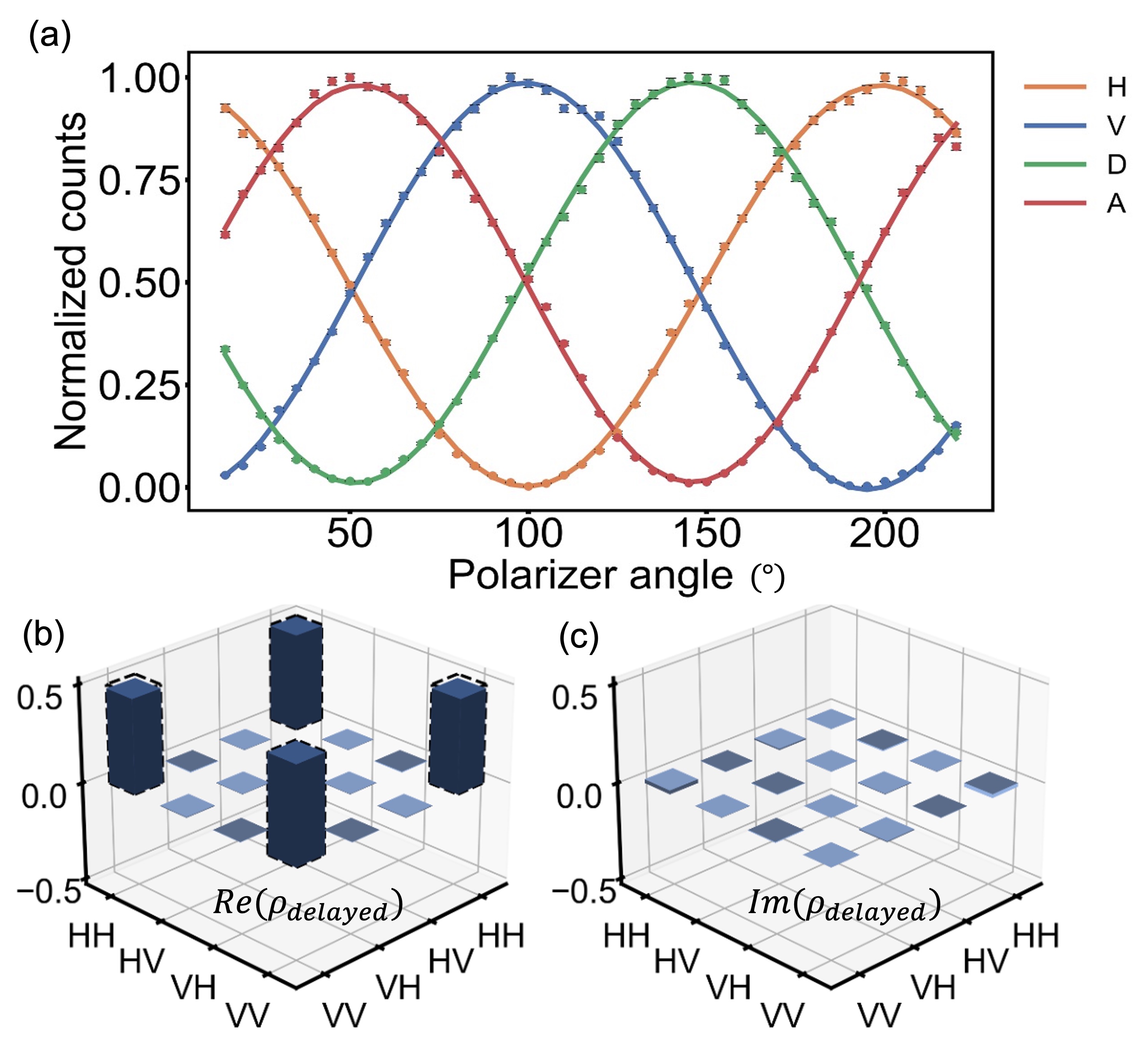}
\caption{
Polarization entanglement characterization after propagation through the optical delay line.
(a) Polarization correlation fringes measured in the H/V and D/A bases. Symbols represent measured normalized coincidence counts, and the solid curves are sinusoidal fits.
(b) Real part of the reconstructed two-photon polarization density matrix for photons after the delay line. Dashed boxes indicate the structure of a maximally entangled polarization state.
(c) Imaginary part of the reconstructed density matrix after the delay line.
The dominant diagonal elements and strong coherence terms between $|HH\rangle$ and $|VV\rangle$ are consistent with the preservation of a polarization entangled state.
}
\label{fig:Polarization}
\end{figure}

Fig.~\ref{fig:EnergyHistogram}(a) and (b) show the time-bin-resolved coincidence histograms for photons measured before and after propagation through the delay line, respectively, under constructive interference conditions. The corresponding histograms under destructive interference are shown in the insets. A pronounced central peak is observed in both cases of constructive interference, arising from the temporal overlap of indistinguishable short-short and long-long two-photon amplitudes.

The coincidence measurements are performed using two silicon avalanche photodiodes (APDs), and the coincidence peak has a fitted FWHM of 567(3) ps, primarily determined by the combined timing jitter. The histograms are recorded with a bin width of 100 ps, and coincidence events are post-selected within a 700 ps window centered on the central peak to isolate the interference contribution \cite{kwiat1993high}, as indicated by the shaded regions in Fig.~\ref{fig:EnergyHistogram}(a) and~(b).

The normalized coincidence counts as a function of the two-photon phase shift $\Delta \varphi$ are shown in Fig.~\ref{fig:EnergyHistogram}(c). Each step corresponds to a mirror displacement of approximately $x=14\,\mathrm{nm}$, which gives $\Delta \varphi \approx 0.43\,\mathrm{rad}$ according to Eq.~(4). The fixed phase offset between the two fringes is attributed to different optical phases accumulated before the analysis interferometer, while energy-time coherence is quantified by the fringe visibility.

High visibility interference fringes are observed for photons measured before and after propagation through the delay line. The measured fringe visibilities are 94.3(2)\% before the delay line and 93.9(3)\% after the delay line. Energy-time interference is further measured after the delay line under two complementary polarization projections, yielding visibilities of $93.9(3)\%$ and $93.8(5)\%$ for $H_{s}H_{i}$ and $D_{s}D_{i}$, respectively. The visibilities under the two projections are consistent within measurement uncertainty. A polarization-dependent change in the energy-time visibility would indicate coupling between the two DOFs, but no such dependence is observed. These joint measurements across both DOFs show no evidence that the delay line introduces cross-correlations, consistent with the preservation of the hyperentangled state.

The measured visibility is mainly limited by the contrast of the analysis interferometer. In the configuration used for Fig.~\ref{fig:EnergyHistogram}(c), the nondegenerate signal and idler photons are analyzed in a common unbalanced interferometer. Optimizing the same interferometer for both wavelengths limits the achievable contrast to 95.1(1)\%, which sets an upper bound on the observed energy-time interference visibility.

The negligible reduction in energy-time visibility indicates that the delay line introduces minimal decoherence to the energy-~time DOF. Using the visibility-based estimate $F_{E} = \frac{1+V}{2}$  \cite{du2025entanglement,steinlechner2017distribution}, the corresponding fidelities are 97.2(1)\% before the delay line and 97.0(2)\% after propagation through the delay line.

Polarization preservation is characterized using the analyzer shown in Fig.~\ref{fig:Setup}(c). Fiber-induced polarization rotations are corrected by polarization compensation modules implemented with quarter-half-quarter wave plate sets. The signal and idler photons are then projected onto selected polarization bases to compare the polarization correlations before and after the FSODL.

Fig.~\ref{fig:Polarization}(a) shows the polarization correlations measured after propagation through the delay line in the H/V and D/A bases. The average visibility is 97.6(1)\% after the delay line, comparable to the reference value of 97.7(2)\% measured before the delay line. The high visibilities in two complementary bases show that the polarization correlations remain robust after the delay line.

Quantum state tomography is performed on the polarization-~entangled photon pairs measured before and after propagation through the delay line using a complete set of polarization projection measurements in the H/V, D/A, and R/L bases. The density matrices $\rho_{\mathrm{nodelay}}$ and $\rho_{\mathrm{delayed}}$ are reconstructed by maximum-likelihood estimation \cite{james2001measurement,altepeter2005photonic}. The fidelity between the reconstructed density matrices is $99.5(4)\%$, evaluated as
\begin{equation}
F_{P} = \left( \mathrm{Tr} \left[ \sqrt{ \sqrt{\rho_{\mathrm{nodelay}}} \, \rho_{\mathrm{delayed}} \, \sqrt{\rho_{\mathrm{nodelay}}} } \right] \right)^2.
\end{equation}

The nonlocal correlations are characterized by directly measuring the Clauser--Horne--Shimony--Holt (CHSH) parameter~$\mathcal{S}$~\cite{clauser1969proposed}. The measured values are $\mathcal{S}= 2.760(9)$ before the delay line and $\mathcal{S}= 2.758(5)$ after the delay line. The comparable Bell violations before and after the delay line demonstrate that the nonlocal polarization correlations are maintained.

%The combined measurements show that the defining polarization and energy-time correlations of the generated hyperentangled photon pairs are preserved after propagation through the FSODL. The energy-time interference confirms that two-photon coherence is maintained after the delay, while the polarization measurements show that the entangled state structure and Bell nonclassicality are retained. These results demonstrate that the FSODL can preserve the tested quantum correlations within a single optical delay line.

The preservation of both DOFs is consistent with the all-reflective design, which suppresses chromatic dispersion and minimizes polarization-dependent disturbances over a broad bandwidth. The preserved energy-time interference indicates that temporal coherence can survive propagation through the delay line, with direct implications for time-bin encoding. In particular, the low dispersion of the FSODL is important for high dimensional time-bin encoding, where multiple temporal modes must maintain their relative timing and phase coherence over the delay~\cite{yu2025quantum}. The compatibility with both temporal coherence and polarization entanglement therefore makes the FSODL suitable for complex photonic states requiring multi-DOF preservation.

Building on this capability, the presented FSODL could be extended into an actively controlled delay-based memory. Detection of the idler photon would herald the presence of its signal partner. The resulting electrical heralding signal could trigger a low-loss, high-speed optical switch~\cite{wang2024low,wang2025low}. The switch would route the signal photon into and out of the recirculating delay line, enabling controlled loading and on-demand retrieval. In this configuration, the storage time could be extended to the microsecond regime, with the required switching architecture and loss budget analyzed in Ref.~\cite{guo2026highly}. Any additional degradation in such a memory would be expected to arise mainly from polarization-dependent switching errors and phase instability, as the FSODL itself has been shown to preserve the hyperentangled state. Such a memory could synchronize independently generated photons for entanglement swapping and quantum repeater protocols~\cite{Briegel1998PRL,Sangouard2011RMP,simon2007quantum}. It could also enable temporal multiplexing, short-term buffering, and adaptive routing of multi-DOF photonic states in scalable quantum networks.

\begin{comment}

\end{comment}

\begin{backmatter}
\bmsection{Funding} %This project is supported by the National Research Foundation, Singapore, through the National Quantum Office, hosted in A*STAR, under its Center for Quantum Technologies Funding Initiative (Grant No. S24Q2d0009), and the Ministry of Education, Singapore, under its Tier 3 grant SENIOR (Grant No. MOE-MOET32024-0009).
Ministry of Education (MOE-MOET32024-0009).

\bmsection{Acknowledgment} 
The authors thank Kwong Chang Jian and Emilien Lavie for comments on the manuscript. The authors also acknowledge the support of Thales Research and Technology Singapore for one co-author’s participation in the experiments.

\bmsection{Disclosures} The authors declare no conflicts of interest.

\bmsection{Data availability} Data underlying the results presented in this paper are not publicly available at this time but may be obtained from the authors upon reasonable request.

\end{backmatter}

%\section{References}

% Bibliography
\bibliography{sample}

% Full bibliography added automatically for Optics Letters submissions; the following line will simply be ignored if submitting to other journals.
% Note that this extra page will not count against page length
\bibliographyfullrefs{sample}

%Manual citation list
%\begin{thebibliography}{1}
%\bibitem{Zhang:14}
%Y.~Zhang, S.~Qiao, L.~Sun, Q.~W. Shi, W.~Huang, %L.~Li, and Z.~Yang,
 % \enquote{Photoinduced active terahertz metamaterials with nanostructured
  %vanadium dioxide film deposited by sol-gel method,} Opt. Express \textbf{22},
  %11070--11078 (2014).
%\end{thebibliography}

% Please include bios and photos of all authors for aop articles
\ifthenelse{\equal{\journalref}{aop}}{%
\section*{Author Biographies}
\begingroup
\setlength\intextsep{0pt}
\begin{minipage}[t][6.3cm][t]{1.0\textwidth} % Adjust height [6.3cm] as required for separation of bio photos.
  \begin{wrapfigure}{L}{0.25\textwidth}
    \includegraphics[width=0.25\textwidth]{john_smith.eps}
  \end{wrapfigure}
  \noindent
  {\bfseries John Smith} received his BSc (Mathematics) in 2000 from The University of Maryland. His research interests include lasers and optics.
\end{minipage}
\begin{minipage}{1.0\textwidth}
  \begin{wrapfigure}{L}{0.25\textwidth}
    \includegraphics[width=0.25\textwidth]{alice_smith.eps}
  \end{wrapfigure}
  \noindent
  {\bfseries Alice Smith} also received her BSc (Mathematics) in 2000 from The University of Maryland. Her research interests also include lasers and optics.
\end{minipage}
\endgroup
}{}

\end{document}